\documentstyle[preprint,aps,pra]{revtex}
\title{Global selection rule in chemical coupling}
\author{Asoke P. Chattopadhyay$^1$ and Anjan K. Dasgupta$^2$\\
$^1$ Department of Chemistry, University of Kalyani\\
Kalyani  741235, India\\
$^2$ Department of Biochemistry, University of Calcutta\\
35 Ballygunje Circular Road, Calcutta  700019, India}
\begin{document}
\maketitle
\vskip 0.8cm
\begin{abstract}
Coupling and decoupling of chemical reactions are explored through a 
modified heat balance equation. Reaction enthalpies are found to play 
crucial role; the sign of their product for a pair of consecutive chemical
reactions determine whether they couple or not. The possibility of a
coupling-uncoupling transition for such reactions is thus introduced for
the first time. The present work resolves a paradox concerning negative
efficiency of coupled consecutive chemical reactions. Enthalpy is also
shown to be a "constant of motion" along the reaction coordinate as long
as the mass action ratio varies little with temperature. The present
analysis puts the observed difference between calorimetric and van't Hoff
enthalpies on a quantitative basis. A case study is presented with a third
order reaction where the enthalpic criterion is useful in selecting the
more probable of the alternative mechanisms.
\end{abstract}
\vskip 1.0cm
PACS No.   82.60.-s 65.40.G
\newpage
\section{Introduction}
The present work extends an earlier attempt\cite{Anjan} to generalize the
conventional phenomenology\cite{Onsager1,Onsager2,Prigogine} for describing
thermodynamics of chemical reactions. The conventional approach, though
based on the limiting assumption of near equilibrium, was applied to
complex processes\cite{Rotten} successfully. Coupling of chemical reactions,
however was treated\cite{Prigogine} using a rather special example of a
cyclic reaction system. For non-cyclic e.g. consecutive reaction schemes,
the Onsager matrix is shown to be diagonal, ruling out phenomenological
coupling. Conventional chemical wisdom, on the other hand, assumes a set of
reactions to be coupled provided they have common intermediate(s). More
importantly, the efficiency of coupling can be shown to be always negative
for coupling schemes treated in the traditional way, whether in linear or
n nonlinear domain. This renders the description of coupling itself as
invalid in the established format. We show, on the contrary, 
how minor modifications in the balance equations lead to alteration in the
definition of macroscopic fluxes and forces in chemical reaction
systems\cite{Shear,Katchal} permitting coupling. The reactions may now be
coupled if enthalpy change in each is non-vanishing. Also, enthalpy
remains a "constant of motion" along reaction coordinate provided the 
mass action ratio has a small variation with temperature. The recent debate
on difference between van't Hoff and calorimetric enthalpies\cite{Grin} 
are put on a more quantitative basis with our approach. A case study
with a third order reaction having two possible mechanisms shows that our
approach correctly identifies the more probable pathway.
\section{The Balance Equation Revisited}
Let us briefly recall the phenomenological equations for chemical coupling
widely used in chemical literature.\cite{Prigogine,Katchal} For a set of
reactions $\rho (=1,....r)$ in a fixed volume, the rates of change of
entropy, concentration etc. are given by\\
\begin{eqnarray}
\partial{s_v}/\partial{t} = -\nabla.J_s + \sigma \\
\partial{c_i}/\partial{t}=-\nabla.J_i+\sum_{\rho} \nu_{i{\rho}}v_{\rho} \\
\partial{q_v}/\partial{t}=-\nabla.J_q+\sum_{\rho}v_{\rho}\Delta H_{\rho} \\
Tds_v = dq_v - \sum_i{\mu_i}dc_i 
\end{eqnarray}
\\
Here, $s_v$, $J_s$ are the density and flux terms for entropy, $c_i$, $J_i$ 
those for the ith chemical species and $q_v$, $J_q$ those for heat. 
$\nu_{i{\rho}}$ is the stoichiometric coefficient of the $i^{th}$ species
in the $\rho^{th}$ chemical reaction for which the velocity is $v_{\rho}$.
$\sigma$ is the entropy production term. While $\Delta H_{\rho}$ is the
enthalpy change of the $\rho^{th}$ chemical reaction, the corresponding Gibbs
free energy change, $\Delta G_{\rho}$, is given by the law of mass action as\\ 
\begin{equation}
\Delta G_{\rho} = -RT ln (v_{\rho}^{+}/v_{\rho}^{-})
\end{equation}
\\
where $v_{\rho}^{+}$ and $v_{\rho}^{-}$ are the forward and reverse reaction
velocities of the $\rho^{th}$ recation, 
and $v_\rho = v_{\rho}^{+}-v_{\rho}^{-}$. From eqs. (1) - (4), equating
non-gradient parts,\\
\begin{equation}
\sigma=J_q.\nabla(1/T)-\sum_i J_i.\nabla(\mu_i/T)+\sum_{\rho}v_{\rho}(\Delta 
H_{\rho}-\sum_i \mu_i \nu_{i{\rho}})/T
\end{equation}
\par
For an isothermal chemical reaction system in a well-stirred (or
homogeneous) medium we get,\\
\begin{equation}
\sigma = \sum_{\rho} v_{\rho} \Delta S_{\rho}
\end{equation}
\\
from $\Delta G_{\rho}=\Delta H_{\rho} - T\Delta S_{\rho}$ and the second law.
Note that the rate of entropy production is obtained as a stoichiometric 
sum of entropy changes of reaction steps, {\it without invoking 
any assumption of linearity of processes}.
\par
     Our eq. (7) can be compared with the standard one for $\sigma$ found in
chemical literature,\cite{Prigogine,Katchal} {\it viz.} \\
\begin{equation}
T\sigma = -\sum_{\rho}v_{\rho}\Delta G_{\rho}
\end{equation}
\\
Following standard phenomenological notations,\\
\begin{equation}
v_{\rho} = -\sum_{\rho^{'}}L_{\rho\rho^{'}}\Delta G_{\rho^{'}}
\end{equation}
\\
Linearising eq. (5), and using $v_{\rho}= v_{\rho}^{+}-v_{\rho}^{-}$   
with eq. (9), we obtain\\
\begin{equation}
L_{\rho\rho^{'}} = \delta_{\rho\rho^{'}} v_{\rho (eq)}^{-}/RT
\end{equation}
\\
where $\delta_{\rho\rho^{'}}$ is the Kronecker delta. Such a diagonal nature
of L makes coupling between two different reactions impossible. Please note
that coupling between cyclic reactions can still emerge in this treatment,
Onsager's example\cite{Prigogine} being the most famous one.
\par
   This impossibility does not occur in our treatment since unlike eq. (8),
 eq. (7) leads to a different phenomenological equation for the reaction
velocity\\
\begin{equation}
v_\rho = -\sum_{\rho^{'}}L_{\rho\rho^{'}}\Delta S_{\rho^{'}}
\end{equation}
\section{Phenomenological vs. Chemical Coupling}
Coupling of chemical reactions, while gaining in popularity over the past
few decades, have received little attention from theoreticians. The major
contributors have been Prigogine and his
coworkers\cite{Prigogine,Prigogine1}. Perhaps, the importance of coupled
reactions is felt nowhere more than in treatments of biochemical
cycles\cite{Lehninger}. Glycolytic or the basic bioenergetic cycle 
(oxidative phosphorylation) are examples of intricate coupling of
consecutive and cyclic reactions\cite{Rotten}. Yet, quantitative
expression of reaction coupling is absent in existing literature in
these fields.
\par
According to Prigogine\cite{Prigogine}, a pair of reactions with affinities
$A_i$ and $A_j$ and velocities $v_i$ and $v_j$ can couple if signs of 
$A_{i}v_{j}$ and $A_{j}v_{j}$ are different. This means that 
one reaction must drive another for the two reactions to couple. Based 
on this idea, Rottenberg\cite{Rotten} defined efficiency of coupling as\\
\[
\eta = -A_{1}v_{1}/A_{2}v_{2}
\]
\\
where the subscripts 1 and 2 denote the driven and the driving reactions
respectively. However, Hill\cite{Hill} pointed out that\\
\[
A \times v \geq 0
\]
\\
for individual reactions. Thus efficiency of coupling is negative and
coupling of consecutive reactions at least is ruled out. This poses a far
stronger challenge to the conventional theory describing coupling, as an
objection to the problem posed by diagonal nature of the Onsager matrix,
as shown in eq. (10), could be that the treatment there is strictly linear.
Although Hill's derivation is consistent with eq. (5), his conjecture that
coupling of chemical reactions is only possible via common intermediates
and has no phenomenological meaning otherwise, is hardly acceptable. In
fact, Hill never sought to explain why a consecutive reaction pair always
has a negative efficiency. The lacuna was in not appreciating a more
fundamental problem, {\it viz.} the nature of the driving force behind
reaction coupling. This impasse has been resolved by the present work,
where the key steps in this regard were modification of eq. (8) to (7),
and of eq. (9) to (11). From eq. (11) it is clear that in our theory 
the efficiency of coupling need not be always negative. Also, 
the Onsager matrix $L_{\rho\rho^{'}}$ is not necessarily diagonal.
It can thus be used as a proper measure of the coupling strength between two
reactions. The general approach is not only useful in resolving the paradox
stated earlier, it also provides a simple thermodynamic criterion for
coupling in simple reaction systems. For example, if the enthalpy of any
reaction step $\rho$ is zero, that reaction can not couple with any other
reaction step $\rho^{'}$ as $L_{\rho\rho^{'}}$ again becomes diagonal.
\section{Gibbs Relation along Reaction Trajectory: $\Delta H_{cal}$ vs 
$\Delta H_{vH}$}
This is an interface between kinetics and thermodynamics of reaction
processes\cite{Prigogine}. For a reaction system the free energy change 
of the $\rho^{th}$ reaction away from equilibrium can be expressed by\\
\begin{equation}
\Delta G_\rho = \Delta G_\rho^0 + RT ln K_\rho
\end{equation}
\\
where the mass action ratio $K_\rho$ is given by\\
\begin{equation}
K_\rho = \prod_jc_j^{\nu_{i\rho}}
\end{equation}
\\
Writing the entropy change along the $\rho^{th}$ reaction as a stoichiometric
sum of molal entropies, $S_i$\\
\begin{equation}
\Delta S_{\rho} = \sum_i \nu_{i\rho}S_i
\end{equation}
\\
From the relation $\Delta S_{\rho} = -\partial{\Delta G_{\rho}}/\partial{T}$, eq. 
(12) and $\Delta G_{\rho} = \Delta H_{\rho} -T\Delta S_{\rho}$,\\
\begin{equation}
\Delta S_{\rho} = \Delta S_{\rho}^0 -RlnK_{\rho} - R(\partial{lnK_{\rho}}/\partial{lnT})
\end{equation}
\\
A simple consequence of eq. (15) is,\\
\begin{equation}
\Delta H_{\rho}=\Delta H_{\rho}^0-RT(\partial{lnK_{\rho}}/\partial{lnT})
\end{equation}
\\
Let us identify the two $\Delta H$ terms appearing in eq. (16) clearly.
At equilibrium, eq. (16) becomes\\
\begin{equation}
\Delta H_{\rho}^{eq}=\Delta H_{\rho}^0-RT[\partial{lnK_{\rho}}/\partial{lnT}]_{eq}
\end{equation}\\
If the second term on RHS becomes very small, we are left with\\
\begin{equation}
\Delta H_\rho \simeq \Delta H_\rho^0 
\end{equation}
\\
Eq. (18) may be stated as follows: {\it the enthalpy remains approximately a 
constant of motion along the reaction co-ordinate for any reaction} (the
van't Hoff approximation). Note that where it is not so, i.e. where
$\Delta H_\rho$ depends on the reaction coordinate and may even change its
sign from $\Delta H_\rho^0$, a pair of coupled reactions may become
decoupled or {\it vice versa}, as the coupling depends on sign of the
product of the enthalpies of the respective reactions. {\it A
coupling-uncoupling transition is therefore possible for such a set of
chemical reactions}. Much further work is needed in this area.
\par
$\Delta H_{\rho}^{eq}$ is clearly the experimentally measured enthalpy
change in a reaction {\it viz.} $\Delta H_{cal}$, which is conventionally
measured at equilibrium (or near equilibrium) conditions. But the first
term on RHS of eq. (16) or (17) can be identified with the van't Hoff
enthalpy of the reaction {\it viz.} $\Delta H_{vH}$, defined as\\
\begin{equation}
\Delta H_{vH} = \Delta H_{\rho}^0 = RT(\partial{lnK_{eq}}/\partial{lnT})
\end{equation}\\
Based on this discussion and the last two equations, we can write\\
\begin{equation}
\Delta H_{cal}=\Delta H_{vH}-RT[\partial{lnK_{\rho}}/\partial{lnT}]_{eq}
\end{equation}\\
There is a large and growing body of 
evidence\cite{Grin,Holtzer,Sturt1,Lumry,Ragone,Weber1,Weber2,Sturt2,Priv}
(including some unpublished work\cite{Stepanov})
of discrepancy between $\Delta H_{cal}$ and $\Delta H_{vH}$. There is also
an awareness of the importance of the ratio of these 
two values, especially in interpreting biocalorimetric data\cite{Babur}. 
While for simple chemical reactions the ratio is close to unity, for reactions
involving macromolecules e.g. in protein folding, there is clear departure of 
the ratio from unity. In biochemical literature,\cite{Priv,Babur} the 
numerical value of the ratio (which may vary from 0.5 to more than
4\cite{Sturt1}, say) is taken to provide a measure of 
cooperativity of the biochemical reaction (e.g. folding). Our analysis gives 
a clear insight into the difference between the two enthalpy values. The
origin of this difference stems from the difference in temperature 
dependence of equilibrium and non-equilibrium mass action ratios. Let us
also recall that such difference between equilibrium and non-equilibrium
values are accepted naturally for Gibbs free energy, for example, and
the difference considered in terms of a mass action ratio.
\par
We realize that further simplification of eq. (20) may be difficult.
Instead, we use certain experimental data to show that the mass action
ratio, $K_{\rho}$, may have a scaling dependence on the experimental
temperature. Holtzer\cite{Holtzer} estimates that the difference between
the calorimetric and van't Hoff enthalpies may be of the order of 45 calories 
per stoichiometric unit for simple chemical reactions. From eq. (20), this
leads to\\
\begin{equation}
\partial{lnK_\rho}/\partial{lnT} \simeq -45/(300 \times 1.98) = -0.075 
\end{equation}
\\
at room temperature ($300^0$ K). We immediately obtain\\
\begin{equation}
K_\rho \simeq const \times T^{-\alpha}
\end{equation}
where $\alpha \simeq 0.075$ for simple chemical reactions.
\par
\section{Coupling Coefficients in Two Step Reaction - Kinetic Approximation}
Using the dissipation eq. (7), we may express the phenomenological relation
of a pair of reactions by\\
\[
\Delta S_2 = R_{21}v_1 + R_{22}v_2
\]
\\
where $v_1$ and $v_2$ are velocities of the two reactions and $R_{ij}$ 
are the coupling terms. The pre-equilibrium condition\cite{Gallene} is given by\\
\begin{equation}
\Delta G_1 \longrightarrow 0 {\hskip 0.5cm} and {\hskip 0.5cm} v_1 \longrightarrow 0
\end{equation}
\\
If we use the approximate linear relation in such cases\\
\begin{equation}
\Delta S_1 = R_{11}v_1+ R_{12}v_2 \simeq R_{12}v_2 
\end{equation}
\\
$\Delta G_1 \simeq 0$ implies $\Delta H_1 \simeq T\Delta S_1$. Using the 
arguments
of eq. (24), the entropy change in the first reaction is given as\\
\begin{equation}
\Delta S_1 \simeq \Delta H_1^0/T
\end{equation}
\\
Using eq. (24), eq. (25) assumes the form\\
\begin{equation}
\Delta H_1^0 = R^{'}_{12}v_2
\end{equation}
\\
where $R^{'}_{12} = TR_{12}$. The approximate form of the dissipation equation 
can be expressed as\\
\begin{equation}
\sigma = \Delta S_1 v_1 + \Delta S_2 v_2 \simeq \Delta S_2 v_2 \geq 0 
\end{equation}
\\
As $v_2$, velocity of the rate-determining step, is positive both for positive
and negative coupling, the pre-equilibrium approximation should satisfy\\
\begin{equation}
\Delta S_2 \geq 0
\end{equation}
\\
Eqs. (25) - (28) express the nature of thermodynamic coupling that may
exist for kinetic schemes satisfying the pre-equilibrium condition. Eq. (26)
implies that velocity of the rate determining step will be proportional
to the enthalpy change in the pre-equilibrium step. It may be interesting
to note that for positive coupling, the pre-equilibrium step must be 
endothermic and vice versa. If on the other hand the coupling is negative
the relation $\Delta H_1^0 \leq 0$ must be satisfied. The negative coupling
implies that one of the reactions among the pair has a negative dissipative
component and is therefore driven by the other.
\section{A case study with in a simple reaction}
Let us consider a typical chemical reaction\\
\[
2NO + O_{2} \rightarrow 2NO_{2}
\]
\\
Using reported molar enthalpy values\cite{Bent} the overall $\Delta H_0$ of 
this reaction is approximately 27.02 Kcal/mole. The standard mechanism 
(henceforth referred to as mechanism I) cited\cite{Benson} is the following
one:\\
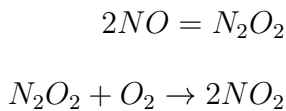
\begin{eqnarray}
2NO = N_2O_2  \nonumber \\
N_{2}O_2 + O_2 \rightarrow 2NO_2 \nonumber
\end{eqnarray}
\\
This mechanism is able to explain the accepted rate law for the overall
reaction {\it viz.} \\
\[
Rate = k[NO]^{2}[O_2]
\]
\\
as also the negative sign of the activation energy of the overall
reaction.\cite{Atkins} Recently,\cite{Plamb} an alternative mechanism 
(referred to as mechanism II), has been proposed, namely\\
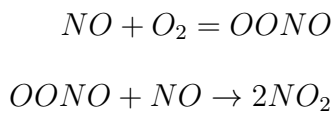
\begin{eqnarray}
NO + O_2 = OONO \nonumber \\
OONO + NO \rightarrow 2NO_2 \nonumber
\end{eqnarray}
\\
which agrees with the rate law given above. According to Plambeck,\cite{Plamb}
spectroscopic evidence admits of simultaneous existence of both these 
mechanisms in the overall reaction, although concentration of OONO may be
larger than $N_2O_2$. Both the intermediates are short lived species, and
although there is speculation about the structure of $N_2O_2$,\cite{Greenw}
nothing is known about OONO.
\par
     We performed {\it ab-initio} calculations on $N_2O_2$ and OONO. The GAMESS
software package\cite{Gam} was used with the ccpVTZ basis set,\cite{Dunn} 
augmented by a d and diffuse s, p type polarization functions. We present
only thermochemical data in Kcal/mol (for $\Delta H^0$ or $\Delta G^0$) or 
cal/mol-K (for $\Delta S^0$). Subscripts 1 and 2 represent the two steps in 
the overall reaction. Along with standard 
data\cite{Bent,Benson,Atkins,Plamb,Greenw} our results are (with an
estimated error limit of $\pm 10$ percent)\\
For mechanism I:\\
\\
$\Delta S_1^0$ = -40.68 cal/mol-K {\hskip 1.8cm} $\Delta S_2^0$ = 5.93 cal/mol-K \\
$\Delta H_1^0$ = -30.2 Kcal/mol {\hskip 2.0cm} $\Delta H_2^0$ = 3.18 Kcal/mol \\
$\Delta G_1^0$ = -17.9 Kcal/mol {\hskip 2.05cm} $\Delta G_2^0$ = 1.4 Kcal/mol \\
\\
Corresponding values for Mechanism II are\\
\\
$\Delta S_1^0$ = -34.34 cal/mol-K {\hskip 1.8cm} $\Delta S_2^0$ = -0.39 cal/mol-k \\
$\Delta H_1^0$ = -11.60 Kcal/mol {\hskip 1.8cm} $\Delta H_2^0$ = -15 Kcal/mol \\
$\Delta G_1^0$ = -1.29 Kcal/mol {\hskip 2.1cm} $\Delta G_2^0$ = -15.29 Kcal/mol \\
\par
It is clear that in both the mechanisms the enthalpy change in the
pre-equilibrium step has a negative sign. Eq. (26) implies that one
reaction must drive the other. From eq. (28) it follows that for both
the mechanisms, $\Delta S_2 = \Delta S_2^0 - RlnK_2 > 0$, where $K_2$ is
the mass action ratio for the second step of the reaction. Therefore,\\
for Mechanism I\\
\begin{equation}
\Delta S_2^0 \geq Rln{[NO_2]^2/([N_2O_2][O_2])} 
\end{equation}
\\
and for Mechanism II\\
\begin{equation}
\Delta S_2^0 \geq Rln{[NO_2]^2/([NO]^2[OONO])} 
\end{equation}
\\
If the intermediate concentration is small, $\Delta S_2^0$ must be greater
than a positive quantity. This need not be the case if the intermediate
has a higher concentration, since then the right hand side of inequality
(30) shifts towards more negative value. Incidentally, the first 
mechanism has a positive $\Delta S_2^0$. In the second mechanism 
$\Delta S_2^0$ has a value approaching zero. In this case, (30) can 
hold good provided the intermediate concentration has a higher value. As 
pointed out by Plambeck,\cite{Plamb} both the mechanisms are known to 
exist, but higher concentration is seen for OONO. The observation is thus 
in accordance with the thermodynamic arguments presented above.
\section{Concluding Remarks}
The present paper shows that coupling of chemical reactions should not be
described in vague qualitative terms e.g. existence of common chemical
intermediates. A pair of reactions remain decoupled as long as the product
of their reaction enthalpies is non-negative. Coupling can only occur if
they have enthalpies of opposite signs. An interesting corollary is that
the same pair of reactions can undergo a transition from coupled to
decoupled state (or {\it vice versa}), provided the mass action ratio
of one or both the reactions change appreciably with progress of the
reaction(s). An important modification brought forward by the present work
is that the rate of internal entropy change near equilibrium is a 
weighted sum of the entropies (and not free energies or chemical 
affinities, as the form in which it is usually expressed)
of the participating reactions. The observed deviation of calorimetric 
(measured) enthalpy from the van't Hoff value for a reaction has also
been explained, the origin of which is shown to be in the difference
in the temperature dependence of equilibrium and non-equilibrium mass
action ratios. For a class of simple chemical reactions, where the deviation
is small, the enthalpy remains a constant of motion along the reaction
coordinate. Finally,
our treatment is shown to identify the more probable of alternate
pathways for a typical third order chemical reaction.\\

\end{document}